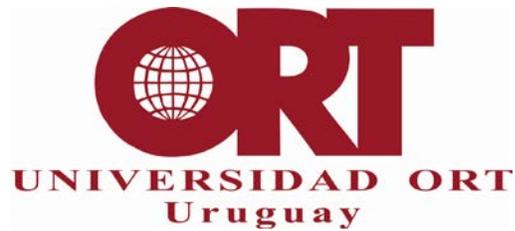

# Opportunities to upgrade the scientific disciplines space.

Gandelman, Néstor[1]
*Universidad ORT Uruguay*

Parcero, Osiris J.[2]
*Kazakh-British Technical University*

Roldán, Flavia[3]♣
*Universidad ORT Uruguay*

Agosto 2021

**Abstract**

Knowledge generated in a given scientific domain may spill over into other close scientific disciplines, thereby improving performance. Using bibliometric data from the SCImago database drawn from a sample of 174 countries, we implement a measure of proximity based on revealed comparative advantage (RCA). Our estimates show that proximity between disciplines positively and significantly affects the RCA growth rate. This impact is larger on disciplines that currently do not have RCA. (JEL codes: L3; L38; O3; O5)

[1] Universidad ORT Uruguay (gandelman@ort.edu.uy)
[2] Kazakh-British Technical University (osiris.jorge.parcero@gmail.com)
[3] Universidad ORT Uruguay (roldan@ort.edu.uy)

♣ The authors acknowledge the financial support of the Agencia Nacional de Investigación e Innovación (ANII), Uruguay, under grant FMV_1_2019_1_156201.
We would like to thank Matilde Pereira for her excellent research assistance. We also thank Almat Kenen, Yerkezhan Kenzheali, and Zhibek Kassymkanova for their invaluable help with data processing.

# I. Introduction

"…science allows us to build taller and taller ladders to reach ever-higher-hanging fruit." In *Building Taller Ladders.* Joel Mokyr, 2018.

In idea-based growth models, economic growth arises from people creating ideas. Ideas are the heart of science, and science, as a source of knowledge, is an engine that drives growth and productivity. This notion can be traced back to Adam Smith (1776) but has also sparked more recent studies aimed at explaining the relationship between scientific research and its economic growth effects.[4] Accordingly, the following questions become relevant. How can a country acquire capabilities that will allow it to produce science? How can a country launch into scientific areas that it has not yet developed? How might a country upgrade its performance in scientific production? This article seeks to address these questions.

In the trade literature, Hidalgo, Klinger, Barabási & Hausmann (2007) (HKBH hereafter) have developed a framework where a country's existing industrial structure determines its potential for technological upgrades and economic development. They find that the proximity between goods already produced with advantages, and those that are not yet competitively produced, plays a crucial role in the success of industrial upgrade processes.

Building on HKBH's (2007) framework, this article examines whether proximity between scientific disciplines affects competitive production. Our main idea is that some set of capabilities, institutions, knowledge, and other inputs necessary to produce science in certain disciplines might be used to produce science in other domains. More precisely, suppose that a country has the advantage of producing science in a particular domain that requires certain inputs. To the extent that these inputs are not completely specific, they might generate advantages (or spillover) in producing advances in other "close" scientific domains, i.e., those requiring similar resources. Consequently, the ability to upgrade a country's science production will depend on how similar (how close) new scientific domains are to those already mastered.

---
[4] Mansfieled (1972 & 1995); Rosenberg (1974); Jaffe (1989); Adams (1990); Rosenberg and Nelson (1994); Partha and David (1994); Stephan (1996); Griliches (1998); Henderson et al. (1998), and even more recently Bloom et al. (2020).

Our methodological challenge is to define closeness between disciplines. A way to address this problem is to observe scientific production patterns. In the spirit of Balassa (1965) for trade patterns and following HKBH (2007), we compute a measure where bibliometric data (publications and citations) reveal proximity among scientific disciplines. The revealed proximity measure indicates the likelihood of taking advantage of existing knowledge or other non-specific inputs with public goods' characteristics that will allow countries to advance in scientific domains, particularly those where the countries currently lack comparative advantage.

To test how proximity between scientific disciplines affects a country's scientific performance, we use the SCImago dataset of world scientific production between 1996 and 2019. Our research offers three main contributions. First, inspired by the trade literature, we propose and implement a measure of proximity between scientific disciplines based on revealed comparative advantage (RCA). Second, we provide evidence that proximity between disciplines positively and significantly affects a country's RCA growth rate. Third, we include a wide range of countries and disciplines from all scientific areas, including Social Science, which has been excluded from most bibliometric studies (Harzing, 2013).

**Place in the literature**

Our article has relevance for several strands of literature. First, it relates to international trade literature, from which we borrow two key and interrelated concepts: revealed comparative advantage and revealed proximity. In particular, by using the RCA measure from Balassa (1965), we compute the revealed proximity index from HKBH (2007). They argue that the more a given product has nearby products, the faster a producing nation can transform its productive profile. In the same vein, articles like Hausmann & Klinger (2006) and Thomson & Athukorala (2020), among others, explore the underlying idea that a country's existing industrial structure determines its industrial upgrade opportunities. All these studies use data on exports from the manufacturing sector. Our paper explores this concept but focuses on scientific knowledge production, proxied by bibliometric data.

The RCA index has been adopted for a wide variety of industries[5] and contexts, including scientometrics (Chuang et al., 2010). The RCA index has many advantages when applied to bibliometric data. Since knowledge is a critical factor in innovation, the study of bibliometric data to assess the state of scientific literature helps to illuminate the scientific capacity of a given nation (Chuang et al., 2010; Radosevic and Yoruk, 2014). In this regard, we place our article in the economics of science literature, as one of its aims is to explain the impact of science on economic growth and productivity (Partha & David, 1994 and Stephan, 1996). Within this literature, a line of research studies linkages between basic science, technological innovation, and economic growth. To cite some relevant studies, Adams (1990) tests the effects of accumulated academic science on productivity in manufacturing industries. Academic science has spillover across industries, and, with long lags (roughly 30 years), it affects industrial productivity. Fleming & Sorenson (2004) study the mechanism through which science accelerates the rate of invention. They argue that science changes inventors' search processes, and they support this claim with an empirical test using patent data. Sorenson & Fleming (2004) consider that the act of publication, which makes information publicly available and encourages the rapid diffusion of knowledge, accounts for the linkage between science and economic growth. They find evidence that publication is an important mechanism in accelerating the rate of technological innovation.

Assuming that science has an impact on economic growth, we investigate factors that boost competitiveness in scientific production and stimulate performance in scientific domains. We claim that there are spillovers between scientific disciplines and that such spillovers occur between close disciplines. Unlike previous articles, which study basic science and patent data, we use bibliometric data for 307 disciplines from Life Sciences, Health Sciences, Physical Sciences, Social Sciences and Humanities, and multidisciplinary fields.

---

[5] For example, forestry (Dieter and Englert, 2007), the manufacture of pharmaceutical products (Cai, 2018), or agriculture and food (Jambor and Babu, 2016). It has also been used in patent analysis (Soete and Wyatt, 1983; Zheng et al., 2011), electronic commerce in the tourism sector and prevalence of the internet (Ruiz Gómez et al., 2018), and start-ups and venture capital (Guerini and Tenca, 2018).

Close to our article, Crespi & Geuna (2008) are among the first to study the relationship between investment in science and research outputs. They explore two characteristics of scientific production: the lag structure of scientific output and knowledge spillovers. Using panel data for a sample of OECD countries, they find that the lag between investment and research output (proxied by publications and citations) is considerable. Also, they quantify spillover of science investment from one country to another using a measure of proximity based on international scientific co-authorship. The higher the level of co-authorship, the more likely the existence of scientific investment spillover between countries. In the same spirit, we use a measure of proximity between scientific disciplines based on their level of co-production, i.e., when a country produces simultaneously in a pair of disciplines. The higher the level of co-production in two scientific domains, the greater the probability of knowledge spillover between disciplines.

Another related study is Davies et al. (2018), which focuses on two neighboring disciplines that share interests and concepts. The authors use case study methodology to examine when sharing promotes cross-fertilization and when it does not. They find that a common language, considered a common input, is essential for knowledge exchange and learning.

The scientometric literature is also interested in the impact of science on the wealth of nations. We can trace a path in the literature from May's (1997) seminal paper on scientific production and national wealth to the present[6].

Cimini et al. (2014) study whether nations tend to specialize or diversify their scientific research and which of these tactics is more efficient in terms of scientific competitiveness. Harzing & Giroud (2014), analyzing a wide range of countries and focus areas, apply the RCA to academic disciplines. They cluster countries and disciplines to explain given countries' scientific profiles and competitiveness.

---

[6] May (1997) compares the top 15 countries, ranked according to the contributions of their scientists to international scientific production. The study covers the period from 1981 to 1994. King (2004) presents a similar analysis to that of May (1997) but using data from 1993-2002. He builds a country ranking according to shares of citations. The study is based on data from more developed nations. In the same strand of literature, we find Laverde-Rojas & Correa (2019), who study the impact of scientific productivity on economic complexity.

Our article shares the general goal of those mentioned above, which is to identify factors contributing to scientific competitiveness. We depart from existing studies in that as we use the revealed proximity measure as a driver in a country's comparative advantage variation rate. This measure allows us to test empirically whether the country's existing scientific profile determines its upgrade opportunities.

The paper is organized as follows. In Section II, we present data and discuss methodology. Section III describes the econometrics and results. We conclude in Section IV.

## II. Data and methodology

This section describes the database and two fundamental concepts underlying the estimated equation presented in Section III. The key concepts are: (i) revealed comparative advantage, and (ii) revealed proximity between disciplines. We use both notions to compute the main variables in the econometric model.

### II.a. Data source

We use public data from the SCImago Journal & Country Rank (SJCR) website based on Scopus. This source covers citable documents per country, including those in over 23,452 peer-reviewed journals (and other serial publications). Our scientific production measure includes citable documents from journals and trade journals (articles, reviews of scientific relevance, and conference papers published in journals). We do not include book series or book reviews, letters, conference meeting abstracts, or non-serial sources. Articles are classified in 307 non-exclusive disciplines. In addition, each paper is non-exclusively classified by the countries of the authors' affiliations. We access data from 1996 to 2019 and include 174 countries that have had at least 100 documents published in 2019.

### II.b. Revealed comparative advantage (RCA)

The direct measure of comparative advantage is a complex calculation because it requires measuring the opportunity costs of production factors. Balassa (1965) developed an indicator that demonstrates how trade patterns reveal which products a country has a comparative advantage.

This indicator has been adopted in a wide variety of contexts, including scientometrics. Formally, the RCA index is defined as the ratio between a discipline's participation in a country's scientific production and the participation of this same discipline in world scientific production. A country has RCA in a particular discipline if within-country participation is larger than expected based on the participation of the discipline in the world scientific production. As proxies for scientific production, we use the number of published documents and citations generated by these documents. Published documents serve as proxies for the quantity of knowledge produced, while citations serve as proxies for its impact. Formally, we calculate RCA for each of the 307 scientific disciplines identified for each country at a given time. The RCA of discipline *i* in country *c* at time *t* is computed as:

$$RCA_{i,t}^c = \frac{x_{i,t}^c / x_t^c}{X_{i,t}^* / X_t^*},$$

where $x_{i,t}^c$ is the number of published documents (citations) of discipline *i* in country *c* at time *t*, $x_t^c$ is the number of documents published (citations) in *all* disciplines in country *c* at time *t*, $X_{i,t}^*$ is the number of documents published (citations) by discipline *i* in the world at time *t*, and $X_t^*$ is the number of documents (citations) published by *all* disciplines in the world at time *t*.

When $RCA_{i,t}^c$ exceeds unity, country *c* at time *t* has a revealed comparative advantage in discipline *i*. Conversely, if $RCA_{i,t}^c$ is less than unity, country *c* has comparative disadvantage in discipline *i* at time *t*.

It is worth noting that the RCA index allows for comparison across scientific domains (within a specific country) and among countries (within a specific discipline). Moreover, the RCA applied to citations avoids the bias generated by the different growth rates in citations per field when the recent years are included in the analysis.

## II.c. Revealed proximity and characterization of the space of scientific disciplines

In this paper, we study whether existing capabilities to produce science positively contribute to producing science in new scientific domains. The main idea is that if a discipline has a comparative advantage in a country, there is some set of capabilities, institutional resources, environment, knowledge, and other inputs that makes this possible. Thus, if two disciplines are similar in the sense that they would require similar capabilities and similar other resources, then it is likely that if we observe RCA in one of them, we will also observe it in the other.

To measure the proximity between two disciplines, we follow a revealed approach based on RCA. This has the same inspiration as RCA: without ex-ante considerations, we let ex-post data "reveal" how similar scientific disciplines are. In this way, we remain agnostic about factors determining proximity between different scientific disciplines.

Following HKBH (2007), revealed proximity is based on conditional probabilities. Looking at world data, we measure the probability of having RCA in discipline *i* conditional on having RCA in area *j*. Since conditional probabilities are not symmetric, we should also consider the converse: the probability of having RCA in area *j* conditional on having RCA in discipline *i*. The two probabilities are not necessarily equal. Formally, the revealed proximity measure is the minimum between these two statistics:

$$\varphi_{ij} = min\{Pr(RCA_{it}|RCA_{jt}), Pr(RCA_{jt}|RCA_{it})\}$$

As the minimum of two conditional probabilities, this measure lies between 0 and 1; the larger the value, the closer the two disciplines are.[7]

Conditional probability is computed for each year using all the countries studied. By the definition of conditional probability, we have:

---

[7] The symmetric imposed solves a technical problem that arises when few countries have RCA in certain disciplines. As an extreme case, suppose that discipline *j* is only produced with RCA by country *c* . Then, for every other discipline produced competitively in country *c*, $Pr(RCA_{it}|RCA_{jt})$ will be equal to 1. This fact would reflect the particular characteristic of the scientific profile of country *c* rather than similarity between disciplines. By taking the minimum, we overcome such a problem. See, for example, Hausmann & Kinger (2006) and Thomson & Athukorala (2020).

$$\varphi_{ijt} = min\{Pr(RCA_{it}|RCA_{jt}), Pr(RCA_{jt}|RCA_{it})\}$$

$$= min\left\{\frac{\Pr(RCA_{it} \cap RCA_{jt})}{\Pr(RCA_{it})}, \frac{\Pr(RCA_{it} \cap RCA_{jt})}{\Pr(RCA_{jt})}\right\}$$

$$= \frac{Pr(RCA_{it} \cap RCA_{jt})}{max\{Pr(RCA_{it}), Pr(RCA_{jt})\}}$$

$$= \frac{\text{number of countries that have RCA in disciplines } i \text{ and } j \text{ at time } t}{max\ \{\text{number of countries with RCA in discipline } i \text{ at time } t,\ \text{number of countries with RCA in discipline } j \text{ at time } t\}}$$

Proceeding in this way, we obtain the yearly revealed proximity matrix that accounts for each discipline's proximity to each remaining discipline. Table 1 lists disciplines that result in greater and lesser proximity for two scientific domains selected for illustration purposes from the 2019 proximity matrix.

**TABLE 1.** Proximity Measures Between Disciplines

*By Publications*

| Business and International Management | | | | Surgery | | | |
|---|---|---|---|---|---|---|---|
| High proximity | | Low proximity | | High proximity | | Low proximity | |
| Business, Management and Accounting (miscellaneous) | 65.50% | Medical Terminology | 1.92% | Pathology and Forensic Medicine | 51.80% | Computer Science Applications | 2.40% |
| Strategy and Management | 63.50% | Reviews and References (medical) | 1.92% | Clinical Psychology | 50.00% | Critical Care Nursing | 0.00% |
| Management, Monitoring, Policy and Law | 55.80% | Aerospace Engineering | 0.00% | Otorhinolaryngology | 50.00% | Environmental Chemistry | 0.00% |

*By Citations*

| Business and International Management | | | | Surgery | | | |
|---|---|---|---|---|---|---|---|
| High proximity | | Low proximity | | High proximity | | Low proximity | |
| Strategy and Management | 60.32% | Medical Terminology | 1.59% | Anesthesiology and Pain Medicine | 54.24% | Medical Assisting and Transcription | 1.69% |
| Social Sciences (miscellaneous) | 61.97% | Assessment and Diagnosis | 3.17% | Cardiology and Cardiovascular Medicine | 54.24% | Nurse Assisting | 1.69% |
| Management of Technology and Innovation | 71.43% | Emergency Medical Services | 4.76% | Emergency Medicine | 57.63% | Drug Guides | 3.39% |

## III. Econometric analyses

We claim that improvements in a scientific discipline can be brought about by successful performance on the part of nearby disciplines. For instance, if a country does well in period *t* in medicine and agriculture (i.e., both produced with RCA greater than one), it would be expected that a close scientific discipline such as veterinary medicine would have significant potential for improvement. Our hypothesis is tested by the following regression:

$$Growth\ RCA_{j,c,t+1} = \alpha_1\ AvgProximity_{j,c,t} + \alpha_2 RCA_{j,c,t} + \eta_j + \eta_t + \varepsilon_{j,c,t} \quad (3)$$

The dependent variable, $Growth\ RCA_{j,c,t+1}$, is annualized RCA growth (either in publications or citations) in scientific discipline *j*, in country *c*, between period *t* and *t+1*. More precisely, using the RCA level for each discipline *j* in each country *c*, we compute geometric average growth for periods 2000-1996, 2004-2000, 2008-2004, 2012-2008, 2016-2012, and 2019-2016 as follows:

$$Growth\ RCA_{j,c,t+1} = \sqrt[n]{\frac{RCA_{j,c,t+1}}{RCA_{j,c,t}}} - 1,$$

where *n* is the number of years in the considered period.

Average Proximity ($AvgProximity_{j,c,t}$) measures the proximity of discipline *j* to those disciplines that have revealed to be produced with comparative advantage in country *c* at time *t*.[8] This variable is computed from the revealed proximity matrix illustrated in Table 1. Formally, $AvgProximity_{j,c,t}$ is calculated as follows:

$$AvgProximity_{j,c,t} = \frac{\sum_i \varphi_{ijt} I_{RCA_{i,t}}}{\sum_i \varphi_{ijt}} \in [0,1],$$

where $I_{RCA_{i,t}}$ is an indicator variable equal to one when country *c* has RCA in scientific domain *i* at time *t*, and zero otherwise. Thus, the numerator is the sum of the proximity of scientific domain *j* to all other disciplines where country *c* has RCA at time *t*. The denominator sums the proximity of domain *j* to all other disciplines at time *t*. Therefore, $AvgProximity_{j,c,t}$ is interpreted as the percentage of scientific space around discipline *j* that is already produced with comparative advantage. For example, $AvgProximity_{j,c,t} = 0.2$ means that 20% of scientific space around *j* is produced with RCA in country c at time t. The larger the $AvgProximity_{j,c,t}$, the greater the probability of knowledge spillover toward *j* from disciplines already producing competitively in country *c*.

---

[8] HKBH (2007) call this measure as *density*.

As in the growth literature, there may be convergence between disciplines. Those that start with lower values might experience more rapid growth rates than those with larger initial values. In order to control for this possible convergence in RCA levels, we control for the level of revealed comparative advantage that domain *j* has in the previous period. If conditional convergence exists, a discipline with a higher RCA value in a particular period should experience lower growth in RCA in the following one (negative $\alpha_2$).

In some specifications, we include in equation (3) an interaction term to test whether the effect of *AvgProximity* depends on the level of RCA.

Finally, $\eta_j$ is a country-discipline fixed effect, $\eta_t$ is a time fixed effect, and $\varepsilon_{j,c,t}$ corresponds to the error term.

It is natural to conjecture that *AvgProximity* might have different effects for disciplines that currently have or do not have comparative advantages. Consequently, we estimate equation (3) in two subsamples. The first group includes scientific disciplines-countries that at the beginning of a period, time *t*, have RCA less than one. The second group includes scientific disciplines-countries that at the beginning of a period, time *t*, have RCA greater than or equal to one.

Conditional on the inclusion of these control variables, our underlying identification assumption is that shocks affecting the proximity measure are uncorrelated with shocks affecting average RCA growth. This assumption is plausible. Let us consider, for example, a shock that occurs in a particular discipline-country that affects the production of documents (or citations) at a particular moment. This shock will affect average RCA growth in such a discipline, but it will not affect the *AvgProximity* variable. This is because that variable is a function of $\varphi_{ijt}$, which considers all countries' information beyond the country where the shock has taken place, as well as a function of indicator variable, $I_{RCA}$, which refers to all other scientific domains different from such a discipline.

### III.a. Descriptive evidence

Table 2 and Table 3 report descriptive statistics for the variables detailed above for the two subsamples in which we divide the dataset, either for the RCA computed from publications (Table 2) or from citations (Table 3). From these tables, three observations are in order.

First, RCA growth has a different pattern depending on whether a scientific domain has achieved comparative advantage or not. In particular, when RCA is less than unity, the mean growth in RCA is $-0.5\%$ for publications and $2.6\%$ for citations. In contrast, when RCA is greater than or equal to one, the mean growth in RCA is $-15.5\%$ for publications and $-17.7\%$ for citations.

Second, the *AvgProximity* variable has a smaller mean for disciplines that do not have a comparative advantage than for disciplines that do. This observation is valid for RCA computed from publications (0.339 vs. 0.402 ) and RCA computed from citations (0.30 vs. 0.374).

Third, the distribution of RCA values exhibits a larger dispersion when disciplines have RCA levels greater than one compared with the dispersion of the RCA'values for disciplines that do not have RCA. This fact holds for RCA from both publications and citations.

**TABLE 2**

**Summary Statistics.** Publications

At time *t*: RCA < 1

|  | Count | Mean | Sd | Min | Max |
|---|---|---|---|---|---|
| *Growth RCA* | 90,702 | -0.005 | 0.312 | -1.000 | 3.604 |
| *AvgProximity* | 90,702 | 0.339 | 0.098 | .0056 | 0.808 |
| *RCA_value* | 90,702 | 0.555 | 0.260 | 0.004 | 0.999 |

At time *t*: RCA≥ 1

|  | Count | Mean | Sd | Min | Max |
|---|---|---|---|---|---|
| *Growth RCA* | 85,359 | -0.155 | 0.335 | -1.000 | 1.504 |
| *AvgProximity* | 85,359 | 0.402 | 0.122 | 0.022 | 1.000 |
| *RCA_value* | 85,359 | 2.689 | 2.435 | 1.000 | 11.385 |

**TABLE 3**

**Summary Statistics.** Citations

At time *t*: RCA < 1

|  | Count | Mean | Sd | Min | Max |
|---|---|---|---|---|---|
| *Growth RCA* | 97,122 | 0.026 | 0.529 | -1.000 | 7.210 |
| *AvgProximity* | 97,122 | 0.300 | 0.102 | 0.018 | 0.769 |
| *RCA_value* | 97,122 | 0.469 | 0.287 | 0.000 | 0.999 |

At time *t*: RCA≥ 1

|  | Count | Mean | Sd | Min | Max |
|---|---|---|---|---|---|
| *Growth RCA* | 73,389 | -0.177 | 0.354 | -1.000 | 2.197 |
| *AvgProximity* | 73,389 | 0.374 | 0.130 | 0.014 | 1.000 |
| *RCA_value* | 73,389 | 2.953 | 2.863 | 1.000 | 12.605 |

Let us now examine in more detail the behavior of the *AvgProximity* variable. Figure 1 presents the distribution of *AvgProximity* for disciplines that initially (at time t) did not have RCA. From this group of disciplines, we identify those that acquired RCA in the next period (solid green line) and those that remained without RCA (red dashed line). Panel (A) draws the *AvgProximity* variable calculated using information from publications, and panel (B) performs this action using information from citations. As we can see, those disciplines that made a jump and gained RCA between periods had a higher value according to our indicator of proximity than those that failed to make such a jump. This fact is supportive of our claim that knowledge in nearby disciplines that is competitively produced might spill over to affect other scientific domains.

**FIGURE 1.**

Average proximity distribution. Disciplines that at time *t* did not have a competitive advantage

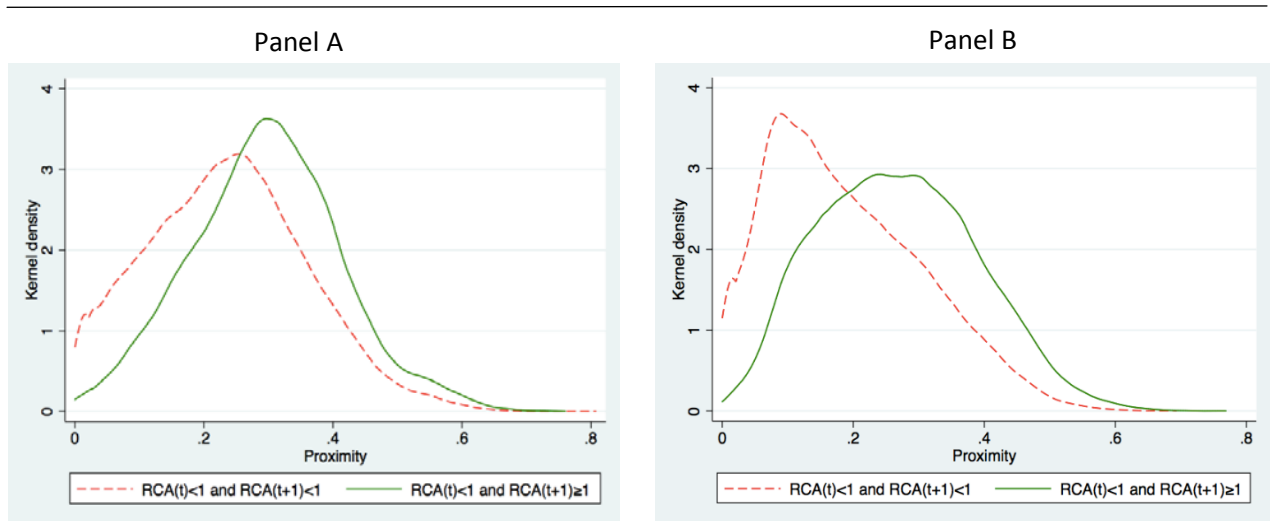

**Figure 1. The distribution of the *AvgProximity* variable.** The graph plots the Kernel density function of the *Avgproximity*$_{j,t}$ variable. Panel (A) draws *Avgproximity* using information from publications, and panel (B) does so using information from citations. The solid green line plots the *Avgproximity* of those disciplines that did not have a comparative advantage but acquired it in the next period. The red dashed line plots the *AvgProximity* of those disciplines that did not have a comparative advantage, and that failed to acquire it in the next period.

Figure 2 describes the distribution of the *AvgProximity* variable for disciplines that initially (at time t) did have RCA, distinguishing two groups: those that maintained RCA in the next period, t+1 (solid blue line), and those that stopped having comparative advantage (black dashed line).

As before, panel (A) draws the *AvgProximity* variable calculated using information from publications, and panel (B) does so using information from citations.

The figures show that *AvgProximity* takes larger values when disciplines maintain their RCA than when disciplines stop having a comparative advantage. This evidence is also consistent with spillover from scientific domains with RCA that help nearby disciplines maintain their competitive status.

**FIGURE 2.**

Average proximity distribution. Disciplines that at time *t* did have a competitive advantage

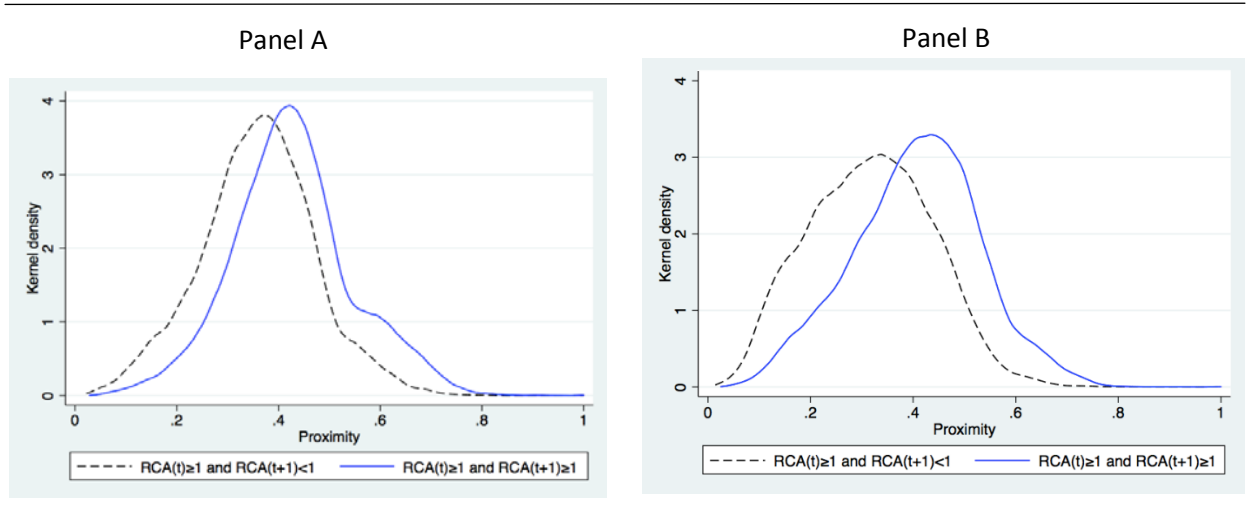

**Figure 2. The distribution of the *AvgProximity* variable.** The graph plots the Kernel density function of the *Avgproximity*$_{j,t}$ variable. Panel (A) draws *Avgproximity* using information from publications, and panel (B) does so using information from citations. The solid blue line plots the *Avgproximity* of those disciplines that did have a comparative advantage at time *t* and maintained it in the next period. The black dashed line plots the *AvgProximity* of those disciplines that did have a comparative advantage at time *t*, but that stopped having it in the next period.

### III.b. Baseline results

We begin with baseline estimates from (3), asking whether the proximity between a given discipline *j* and the set of disciplines that produce competitively in country *c* (with RCA≥1) will affect the rate at which the country's RCA grows. Thus, the critical parameter is the coefficient of *AvgProximity*, coefficient $\alpha_1$.

Table 4.a and Table 4.b report the estimate of the coefficients. The dependent variable is *RCA Growth* in discipline *j* in country *c* at time *t+1*, when we compute RCA for publications (Table

4.a) or for citations (Table 4.b). As indicated earlier, all specifications include time and country-discipline fixed effects.

From both tables, we observe that the impact of the proximity between discipline *j* and the set of disciplines produced competitively in country *c* (the *AvgProximity* variable) has different behavior depending on whether the discipline reveals not to have (Columns 1 and 3) or to have (Columns 2 and 4) comparative advantage.

The first and second columns of Table 4.a and 4.b indicate that the *Avg Proximity* variable has a positive impact on the growth rate of the discipline's RCA, both for publications and citations. Nonetheless, this result hides differences according to the level of RCA that a discipline has. Columns 3 and 4 in Table 4.a and 4.b. report these heterogeneous effects in detail. Accounting for the marginal effect of *AvgProximity*, if a country does not have a comparative advantage, an increase in one standard deviation in *AvgProximity* will push up growth in RCA by 5.65 percentage points for publications[9] and 8.05 for citations. Moreover, when the country already had a comparative advantage in a discipline, the RCA growth rate increases by 5.92 percentage points for publications and 3.61 for citations. As we noted before, this positive effect varies according to the RCA level that the discipline has. Thus, when at time *t*, the discipline has RCA<1, a larger *AvgProximity* will larger increase the RCA's growth rate, as larger the level of the discipline's RCA is. In contrast, when at period *t*, the discipline has RCA $\geq$ 1, *AvgProximity* will increase the RCA's growth rate, but less as larger the discipline's RCA level. This observation holds when we compute RCA using information from either publications or citations, though when RCA<1, the point estimate for publications remains positive but is less precisely estimated.

---

[9] This impact is computed as follows: the marginal effect of *AvgProximity* for publications when RCA<1 (0.577) times one standard deviation of *AvgProximity* in such a case (0.098). The remaining impacts on the growth rates are calculated in the same way.

**TABLE 4.a**

Average Proximity between Disciplines and RCA Growth Rates

| | PUBLICATIONS | | | |
|---|---|---|---|---|
| | (1) | (2) | (3) | (4) |
| VARIABLES | At t:RCA < 1 | At t:RCA≥ 1 | At t:RCA < 1 | At t:RCA≥ 1 |
| *AvgProximity* | 0.580*** | 0.547*** | 0.496*** | 0.605*** |
| | (0.064) | (0.056) | (0.091) | (0.063) |
| *RCA* | -0.611*** | -0.010*** | -0.659*** | -0.036*** |
| | (0.011) | (0.002) | (0.034) | (0.003) |
| *AvgProximity* x *RCA* | | | 0.145 | -0.044*** |
| | | | (0.099) | (0.009) |
| Constant | 0.052** | -0.361*** | 0.079*** | -0.258*** |
| | (0.020) | (0.027) | (0.030) | (0.025) |
| Marginal effects | | | | |
| *AvgProximity* | | | 0.577*** | 0.486*** |
| | | | (0.0649) | (0.0485) |
| *RCA* | | | -0.610*** | -0.0536*** |
| | | | (0.0109) | (0.00134) |
| Observations | 90,702 | 85,359 | 90,702 | 85,359 |
| R-squared | 0.203 | 0.062 | 0.203 | 0.122 |
| Number of Country-Discipline units | 28,234 | 29,426 | 28,234 | 29,426 |
| Fixed Effects | YES | YES | YES | YES |
| Year FE | YES | YES | YES | YES |

Notes: Cluster robust standard errors in parentheses. This table presents the estimate for the average proximity effect on average RCA annual growth in periods 2000-1996, 2004-2000, 2008-2004, 2012-2008, 2016-2012, and 2019-2016, when RCA is computed using information from documents. Regressions in the first and third columns consider country-discipline observations such that at the beginning of the period, RCA is less than 1, while the second and fourth columns consider observations such that at the beginning of the period, RCA is at least equal to 1. Each regression includes country-discipline fixed effects and year dummies. Regressions control for RCA value at the country-discipline-year level at the beginning of the period.
*** Significant at the 1 percent level. **Significant at the 5 percent level. *Significant at the 10 percent level.

**TABLE 4.b**

Average Proximity between Disciplines and RCA Growth Rates

| | CITATIONS | | | |
|---|---|---|---|---|
| | (1) | (2) | (3) | (4) |
| VARIABLES | At t:RCA < 1 | At t:RCA≥ 1 | At t:RCA < 1 | At t:RCA≥ 1 |
| *AvgProximity* | 0.802*** | 0.247*** | 0.348*** | 0.418*** |
| | (0.072) | (0.041) | (0.103) | (0.051) |
| *RCA* | -0.903*** | -0.046*** | -1.180*** | -0.033*** |
| | (0.017) | (0.001) | (0.040) | (0.003) |
| *AvgProximity* x *RCA* | | | 0.939*** | -0.048*** |
| | | | (0.123) | (0.009) |
| Constant | 0.112*** | -0.155*** | 0.237*** | -0.213*** |
| | (0.022) | (0.017) | (0.031) | (0.020) |
| Marginal effects | | | | |
| *AvgProximity* | | | 0.789*** | 0.278*** |
| | | | (0.0753) | (0.0422) |
| *RCA* | | | -0.899*** | -0.0509*** |
| | | | (0.0146) | (0.00138) |
| Observations | 97,122 | 73,389 | 97,122 | 73,389 |
| R-squared | 0.177 | 0.130 | 0.179 | 0.132 |
| Number of Country-Discipline units | 31,846 | 27,307 | 31,846 | 27,307 |
| Fixed Effects | YES | YES | YES | YES |
| Year FE | YES | YES | YES | YES |

Notes: Cluster robust standard errors in parentheses. This table presents the estimate for the average proximity effect on average RCA annual growth in periods 2000-1996, 2004-2000, 2008-2004, 2012-2008, 2016-2012, and 2019-2016, when RCA is computed using information from citations. Regressions in the first and third columns consider country-discipline observations such that at the beginning of the period, RCA is less than 1, while the second and fourth columns consider observations such that at the beginning of the period, RCA is at least equal to 1. Each regression includes country-discipline fixed effects and year dummies. Regressions control for RCA value at the country-discipline-year level at the beginning of the period.
*** Significant at the 1 percent level. **Significant at the 5 percent level.  *Significant at the 10 percent level.

The above estimates confirm the intuition that the RCA growth rate in scientific domains depends on the average proximity that each discipline has to other scientific domains in which a country currently has a comparative advantage.

We now illustrate our results using the estimation reported in Table 4 (only for information from publications). We project the RCA growth rate of each discipline due only to the impact of the AvgProximity variable. Figure 3 plots RCA growth projected for the following years against 2019's average proximity for a small set of countries chosen only to illuminate the intuition of our results.[10]

For China and South Korea, disciplines in the Physical Sciences are closer to other domains in which these countries currently have comparative advantages and have the largest expected growth rates. In contrast, for Uruguay and Ghana, disciplines in the Physical Sciences appear closer to the origin of the graphs, pointing to more isolation of disciplines where these countries have comparative advantages and have lower expected RCA growth rates.

---

[10]For more details, in the Appendix, we include for each of these countries the five disciplines that are likely to grow the most and the five disciplines that are likely to grow the least given this projection when RCA<1.

**FIGURE 3.**

Projection of RCA Growth Rate

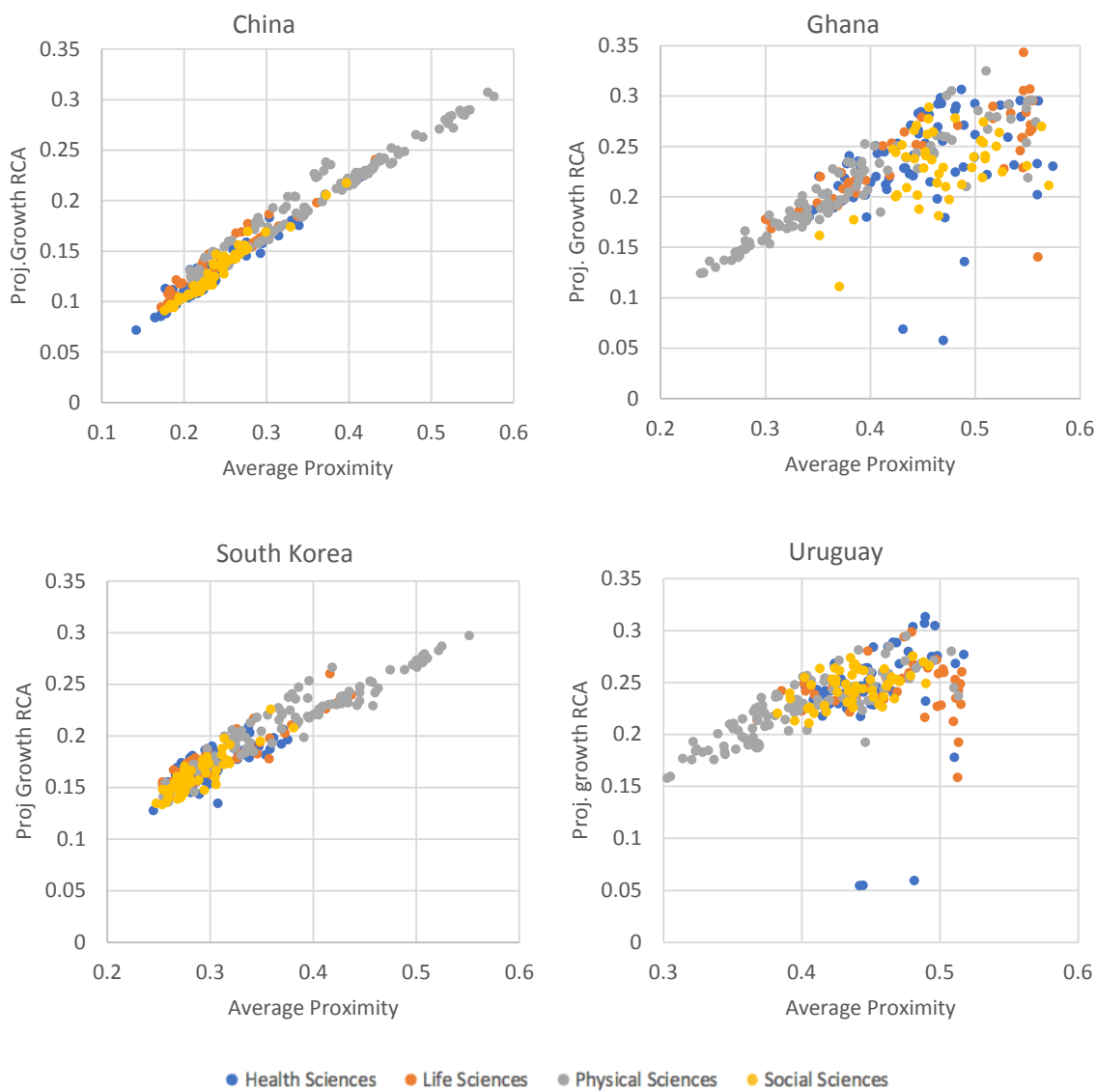

**Figure 3. Projection RCA growth rates.** The graph plots projected RCA growth rates for the following years due to the effect of the average proximity variable using estimates from Table 4.a. Countries are chosen only to illustrate the results.

A significant portion of disciplines with RCA below unity in 2019 but with relatively high growth rates will achieve RCA greater than unity in the next period. From the pictures, we can conclude that the more developed a nearby scientific space (larger average proximity), the higher the RCA growth rate of a specific discipline.

The estimations in Table 4.a and 4.b and the above examples suggest that greater dynamism will be observed in scientific disciplines that exhibit higher average proximity to scientific space that has already been developed.

### III.c. Robustness

We now run the following exercises for robustness,[11] where we use an alternative measure of average proximity and alternative measures of scientific performance.

**An alternative measure of average proximity.** The average proximity variable captures the idea that knowledge and skills, among other variables, accumulated by a country in the set of disciplines that it currently produces with advantage can spill over to nearby disciplines and boost its RCA growth rate.

It is worth noting that given how we compute the *AvgProximity* variable, it correlates with countries' attributes or factors that, simultaneously and competitively, produce pairs of scientific disciplines. A natural concern is whether all these factors are transferable in such a way that they contribute to producing science in nearby disciplines. This concern exists because some attributes play only minor roles in establishing or preserving RCA. Among these factors, we can distinguish (i) factors or attributes like language and natural resources that do not change over time; (ii) factors such as institutional frameworks or policies (related to education and support for science and innovation) that are somewhat stable over time; and (iii) factors like transferable skills, shared resources not exhausted in producing one of the disciplines, knowledge, technologies, and techniques transferable between areas. We claim that this last category is more likely to be directly linked to the production process in science.

Although it is difficult to distinguish among the three categories above, we follow Thomson & Athukorala (2020) and explore the impact of factors that change over time to focus on the third set. We measure the change in the variable *AvgProximity* to isolate these factors, eliminating the impact of factors that do not vary over time. The underlying assumption in this estimation is that noise in the *AvgProximity* variable at time *t+1* is not correlated with its error at time *t*, so it plays no systematic errors in the estimate. In the proximity matrix, given the large number of observations at each period (a matrix of 307 x 307) and the large number of countries in our sample (174), we can safely assume that distortions or shocks that affect a single country or discipline can be smoothed out. Moreover, as we explained earlier, noise in the *AvgProximity* variable is uncorrelated with $\varepsilon_{j,c,t}$ in equation (3).

---

[11] We also run various other exercises (not reported) where we include additional controls, for instance Gross Domestic Product per capita, to capture differences in resources available. The results remain almost identical.

Table 6 reports our results. As can be seen, they are similar and confirm that *AvgProximity* plays a significant role in determining RCA growth rates.

**TABLE 6.** Change in Average Proximity and RCA Growth Rates

|  | PUBLICATIONS | | CITATIONS | |
| --- | --- | --- | --- | --- |
|  | (1) | (2) | (3) | (4) |
| VARIABLES | At t:RCA < 1 | At t:RCA≥ 1 | At t:RCA < 1 | At t:RCA≥ 1 |
| Δ $AvgProximity$ | 0.695*** | 0.826*** | 1.050*** | 0.837*** |
|  | (0.040) | (0.046) | (0.057) | (0.046) |
| $RCA\_j,c,t$ | -0.593*** | -0.053*** | -0.857*** | -0.047*** |
|  | (0.011) | (0.001) | (0.016) | (0.001) |
| Constant | 0.223*** | -0.026*** | 0.308*** | -0.052*** |
|  | (0.006) | (0.005) | (0.008) | (0.005) |
| Observations | 90,702 | 85,359 | 97,122 | 73,389 |
| R-squared | 0.210 | 0.141 | 0.187 | 0.163 |
| Number of Country-Discipline units | 28,234 | 29,426 | 31,846 | 27,307 |
| Fixed Effects | YES | YES | YES | YES |
| Year FE | YES | YES | YES | YES |

Notes: Robust clustered standard errors in parentheses. This table presents the estimate for variation in the average proximity effect on average RCA annual growth in periods 2000-1996, 2004-2000, 2008-2004, 2012-2008, 2016-2012, and 2019-2016, when RCA is computed using information from publications and citations. Regressions in the first and third columns consider country-discipline observations such that at the beginning of the period, RCA is less than 1, while the second and fourth columns consider observations such that at the beginning of the period, RCA is at least equal to 1. Each regression includes country-discipline fixed effects and year dummies. Regressions control for RCA value at the country-discipline-year level at the beginning of the period.
*** Significant at the 1 percent level.   **Significant at the 5 percent level.   *Significant at the 10 percent level.

**Alternative measures of scientific performance.** Using information regarding publications and citations, we compute their annualized growth for each discipline country. Table 7 reports the impact of average proximity on these alternative measures of scientific production. We find that the main results of our baseline specification still hold.

**TABLE 7**

Average Proximity between Disciplines and Growth Rates on Publication and Citations

|  | PUBLICATIONS | | CITATIONS | |
|---|---|---|---|---|
|  | (1) | (2) | (3) | (4) |
| VARIABLES | At t:RCA < 1 | At t:RCA≥ 1 | At t:RCA < 1 | At t:RCA≥ 1 |
| $AvgProximity$ | 0.651*** | 0.594*** | 0.541*** | 0.411*** |
|  | (0.097) | (0.076) | (0.136) | (0.098) |
| $RCA$ | -0.692*** | -0.011*** | -0.744*** | -0.013*** |
|  | (0.014) | (0.002) | (0.023) | (0.003) |
| Constant | 0.149*** | -0.297*** | 0.356*** | -0.176*** |
|  | (0.030) | (0.033) | (0.041) | (0.041) |
| Observations | 88,320 | 83,614 | 66,634 | 65,182 |
| R-squared | 0.196 | 0.063 | 0.100 | 0.049 |
| Number of Country-Discipline units | 27,486 | 28,800 | 23,656 | 25,283 |
| Fixed Effects | YES | YES | YES | YES |
| Year FE | YES | YES | YES | YES |

Notes: Robust clustered standard errors in parentheses. This table presents the estimate for the average proximity effect on annual growth (columns 1 and 2) as revealed by publications and on annual growth as revealed by citations (columns 3 and 4) in periods 2000-1996, 2004-2000, 2008-2004, 2012-2008, 2016-2012, and 2019-2016. Regressions in the first and third columns consider country-discipline observations such that at the beginning of the period, RCA is less than 1, while the second and fourth columns consider observations such that at the beginning of the period, RCA is at least equal to 1. Each regression includes country-discipline fixed effects and year dummies. Regressions control for RCA value at the country-discipline-year level at the beginning of the period.
*** Significant at the 1 percent level.   **Significant at the 5 percent level.   *Significant at the 10 percent level.

## IV. Discussion and Conclusions

This paper has provided evidence of how proximity between scientific disciplines plays a crucial role in the process of scientific upgrades.

The underlying idea is that there is likely to be knowledge spillover between scientific disciplines and that such spillover is larger between close scientific disciplines. We have implemented a definition of closeness according to what the data reveals. Using the SCImago dataset for publications and citations, empirical findings support the notion that the growth rate of comparative advantages in scientific domains depends on the average proximity that each discipline has to those scientific domains where a country currently has a comparative advantage.

These average effects conceal a diverse range of impact across disciplines. In particular, for scientific domains that are not yet producing competitively, greater proximity will have a larger effect as a scientific domain's RCA level increases.

The finding that the initial level of revealed comparative advantage has a negative and statistically significant effect on the growth rate of RCA can be interpreted through the lens of convergence.

These results may have significant policy implications. Considering that there are factors like skills, knowledge, technologies, and techniques transferable between scientific domains, public and private efforts to support bundled disciplines will have increasing returns due to positive externalities over proximate scientific domains.

# REFERENCES

ADAMS, J. D. (1990). Fundamental stocks of knowledge and productivity growth. *Journal of Political Economy*, *98*(4), 673-702.

BLOOM, N., JONES, C. I., VAN REENEN, J. & WEBB, M. (2020). Are ideas getting harder to find? *American Economic Review*, 110(4), 1104-44.

CAI, J., ZHAO, H. & COYTE, P. C. (2018). The effect of intellectual property rights protection on the international competitiveness of the pharmaceutical manufacturing industry in China. *Engineering Economics*, 29(1), 62-71.

CIMINI, G., GABRIELLI, A. & LABINI, F. S. (2014). The scientific competitiveness of nations. *PloS one*, 9(12). Recuperado de https://doi.org/10.1371/journal.pone.0113470

CRESPI, G. A. & GEUNA, A. (2008). An empirical study of scientific production: a cross country analysis, 1981–2002. *Research Policy*, 37(4), 565-579.

DAVIES, A., MANNING, S. & SÖDERLUND, J. (2018). When neighboring disciplines fail to learn from each other: the case of innovation and project management research. *Research Policy*, 47(5), 965-979.

DIETER, M. & ENGLERT, H. (2007). Competitiveness in the global forest industry sector: an empirical study with special emphasis on Germany. *European Journal of Forest Research*, 126(3), 401-412.

FLEMING, L. & SORENSON, O. (2004). Science as a map in technological search. *Strategic Management Journal*, *25*(8-9), 909-928.

GRILICHES, Z. (1998). R & D and productivity: the unfinished business. *Estudios de Economía*, *25*(2), 145-160.

GUERINI, M. & TENCA, F. (2018). The geography of technology-intensive start-ups and venture capital: European evidence. *Economia e Politica Industriale*, 45(3), 361-386.

GUEVARA, M. & MENDOZA, M. (2013). Revealing comparative advantages in the backbone of science. In *CompSci'13: Proceedings of the 2013 Workshop on computational scientometrics: theory & applications* (pp. 31-36). San Francisco, CA: USA.

HARZING, A. W. (2013). A preliminary test of Google Scholar as a source for citation data: a longitudinal study of Nobel prize winners. *Scientometrics*, 94(3), 1057-1075.

HARZING, A. W. & GIROUD, A. (2014). The competitive advantage of nations: an application to academia. *Journal of Informetrics*, 8(1), 29-42.

HAUSMANN, R. & KLINGER, B. (2006). *Structural transformation and patterns of comparative advantage in the product space.* (CID Working Paper nº128). Cambridge; Massachusetts: CID.

HENDERSON, R., JAFFE, A. B. & TRAJTENBERG, M. (1998). Universities as a source of commercial technology: a detailed analysis of university patenting, 1965–1988. *Review of Economics and Statistics*, 80(1), 119-127

HIDALGO, C. A., KLINGER, B., BARABÁSI, A. L., & HAUSMANN, R. (2007). The product space conditions the development of nations. *Science*, 317(5837), 482-487.

JAFFE, A. B. (1989). Real effects of academic research. *The American Economic Review*, 79(5), 957-970.

JAFFE, K., et.al. (2013). Productivity in physical and chemical science predicts the future economic growth of developing countries better than other popular indices. *PloS One*, 8(6). Recuperado de https://doi.org/10.1371/journal.pone.0066239

JAFFE, K., TER HORST, E., GUNN, L. H., ZAMBRANO, J. D., & MOLINA, G. (2020). A network analysis of research productivity by country, discipline, and wealth. *PloS one*, 15(5). Recuperado de https://doi.org/10.1371/journal.pone.0232458

JAMBOR, A. & BABU, S. C. (2017). *Competitiveness of global agriculture: policy lessons for food security.* Washington, D.C: International Food Policy Research Institute. Recuperado de https://doi.org710.2499/9780896292598

KING, D. A. (2004). The scientific impact of nations. *Nature*, 430(6997), 311-316.

MANSFIELD, E. (1972). Contribution of R&D to economic growth in the United States. *Science*, 175(4021), 477-486.

MANSFIELD, E. (1995). Academic research underlying industrial innovations: sources, characteristics, and financing. *The Review of Economics and Statistics*, 77(1), 55-65.

MAY, R. M. (1997). The scientific wealth of nations. *Science*, 275(5301), 793-796.

PARTHA, D. & DAVID, P. A. (1994). Toward a new economics of science. *Research Policy*, 23(5), 487-521.

RADOSEVIC, S. & YORUK, E. (2014). Are there global shifts in the world science base?: analysing the catching up and falling behind of world regions. *Scientometrics*, 101(3), 1897-1924.

ROSENBERG, N. (1990). Why firms do basic research (with their own money)? *Research Policy*, 19(2), 165-174.

ROSENBERG, N. & NELSON, R. R. (1994). American universities and technical advance in industry. *Research Policy*, 23(3), 323-348.

RUIZ GÓMEZ, L. M., RODRÍGUEZ FERNÁNDEZ, L. & NAVIO-MARCO, J. (2018). Application of communication technologies (ICT) within the tourism industry in the European Union. *Tourism: An International Interdisciplinary Journal*, 66(2), 239-245.

SOETE, L. & WYATT, S. (1983). The use of foreign patenting as an internationally comparable science and technology output indicator. *Scientometrics*, 5(1), 31-54.

SORENSON, O. & FLEMING, L. (2004). Science and the diffusion of knowledge. *Research Policy*, 33(10), 1615-1634.

STEPHAN, P. E. (1996). The economics of science. *Journal of Economic Literature*, 34(3), 1199-1235.

THOMSON, R. & ATHUKORALA, P. C. (2020). Global production networks and the evolution of industrial capabilities: does production sharing warp the product space?. *Oxford Economic Papers*, 72(3), 731-747.

ZHENG, J., et.al. (2011). Industry evolution and key technologies in China based on patent analysis. *Scientometrics*, 87(1), 175-188.

ZHOU, P. & LEYDESDORFF, L. (2006). The emergence of China as a leading nation in science. *Research Policy*, 35(1), 83-104.

# APPENDIX

This Appendix illustrates our results in Table 4 using information from China, Ghana, South Korea, and Uruguay. Tables below detail those disciplines that in 2019 did not have an RCA greater than unity (computed from the information of publications). Hence, we can see these disciplines as the most promising ones.

Using the estimation reported in Table 4.a, we project the RCA growth rate of each discipline due to the impact of the *AvgProximity* variable. The tables inform us of the five disciplines that will grow the most and the five disciplines that will grow the least.

**Table A.1.** Projection RCA growth rate of most promising disciplines. China

|  | Projection | RCA 2019 | Main area |
|---|---|---|---|
| **Lower projected growth** | | | |
| Review and Exam Preparation | 7,2% | 0,08 | Health Sciences |
| Leadership and Management | 8,4% | 0,11 | Health Sciences |
| Medical and Surgical Nursing | 8,5% | 0,13 | Health Sciences |
| Care Planning | 8,6% | 0,01 | Health Sciences |
| Critical Care Nursing | 8,7% | 0,07 | Health Sciences |
| **Higher projected growth** | | | |
| Radiation | 22,4% | 0,85 | Physical Sciences |
| Safety, Risk, Reliability and Quality | 22,7% | 0,95 | Physical Sciences |
| Statistical and Nonlinear Physics | 23,0% | 0,87 | Physical Sciences |
| Physics and Astronomy (miscellaneous) | 23,6% | 0,89 | Physical Sciences |
| Computer Science (miscellaneous) | 23,8% | 1,00 | Physical Sciences |

**Table A.2.** Projection RCA growth rate of most promising disciplines. Ghana

|  | Projection | RCA 2019 | Main area |
|---|---|---|---|
| **Lower projected growth** | | | |
| Materials Chemistry | 12.4% | 0.18 | Physical Sciences |
| Catalysis | 12.5% | 0.15 | Physical Sciences |
| Condensed Matter Physics | 12.5% | 0.16 | Physical Sciences |
| Electronic, Optical and Magnetic Materials | 13.0% | 0.14 | Physical Sciences |
| Chemistry (miscellaneous) | 13.6% | 0.39 | Physical Sciences |
| **Higher projected growth** | | | |
| Pharmacology (medical) | 29.3% | 0.62 | Health Sciences |
| Family Practice | 29.3% | 0.90 | Health Sciences |
| Emergency Medicine | 29.8% | 0.99 | Health Sciences |
| Geochemistry and Petrology | 30.1% | 0.97 | Physical Sciences |
| Geology | 30.5% | 0.99 | Physical Sciences |

**Table A.3.** Projection RCA growth rate of most promising disciplines. South Korea

|  | Projection | RCA 2019 | Main area |
|---|---|---|---|
| **Lower projected growth** | | | |
| Review and Exam Preparation | 12.8% | 0.18 | Health Sciences |
| Anthropology | 13.3% | 0.22 | Social Sciences |
| History and Philosophy of Science | 13.5% | 0.33 | Social Sciences |
| Demography | 13.6% | 0.22 | Social Sciences |
| Nature and Landscape Conservation | 13.6% | 0.20 | Physical Sciences |
| **Higher projected growth** | | | |
| Artificial Intelligence | 24.1% | 0.96 | Physical Sciences |
| Computer Networks and Communications | 24.7% | 1.00 | Physical Sciences |
| Colloid and Surface Chemistry | 25.3% | 1.00 | Physical Sciences |
| Structural Biology | 26.0% | 0.90 | Life Sciences |
| Signal Processing | 26.7% | 0.97 | Physical Sciences |

**Table A.4.** Projection RCA growth rate of most promising disciplines. Uruguay

|  | Projection | RCA 2019 | Main area |
|---|---|---|---|
| **Lower projected growth** | | | |
| Mechanics of Materials | 15.8% | 0.18 | Physical Sciences |
| Mechanical Engineering | 16.0% | 0.19 | Physical Sciences |
| Ceramics and Composites | 17.6% | 0.36 | Physical Sciences |
| Polymers and Plastics | 17.6% | 0.18 | Physical Sciences |
| Materials Science (miscellaneous) | 17.7% | 0.46 | Physical Sciences |
| **Higher projected growth** | | | |
| Water Science and Technology | 29.5% | 0.85 | Physical Sciences |
| Pharmacology, Toxicology and Pharmaceutics (miscellaneous) | 29.9% | 0.87 | Life Sciences |
| Pediatrics, Perinatology and Child Health | 30.4% | 0.94 | Health Sciences |
| Public Health, Environmental and Occupational Health | 30.5% | 0.81 | Health Sciences |
| Epidemiology | 30.7% | 0.91 | Health Sciences |